\documentclass[12pt]{article}
\usepackage{amsmath}
\usepackage{graphicx,psfrag,epsf}
\usepackage{natbib}
\usepackage{url} 

\usepackage{setspace}
\usepackage{color,soul}
\usepackage{mathrsfs}   %$\mathscr{A}$

\newcommand{\blind}{0}

\begin{document}	
\if0\blind
{
	\title{\bf Asymmetric Dependence Measurement and Testing}
\author{ H. D. Vinod 
	\thanks{address: H. D. Vinod, 
		Professor of Economics, Fordham University,
		Bronx, New York, USA 10458. 
		E-mail: vinod@fordham.edu. 
		JEL codes C30, C51. 
}} 
	\maketitle
} \fi
\if1\blind
{
	\bigskip
	\bigskip
	\bigskip
	\begin{center}
		{\LARGE\bf Asymmetric Dependence Measurement}
	\end{center}
	\medskip
} \fi

%	\maketitle
	
\begin{abstract}
Measuring the (causal) direction and strength
of dependence between two variables (events),
$X_i$ and $X_j$, is fundamental for all science. 
Our survey of decades-long literature on statistical dependence 
reveals that most assume symmetry in the sense that
the strength of dependence of
$X_i$ on $X_j$ exactly equals
the strength of dependence of
$X_j$ on $X_i$. However, we
show that such symmetry is often untrue in
many real-world examples, being
neither necessary nor sufficient. 
Vinod's (2014) asymmetric matrix $R^*\in [-1,1]$ of generalized correlation
coefficients provides intuitively appealing, readily interpretable, and 
superior measures of dependence. This paper
proposes statistical inference 
for $R^*$ using Taraldsen's (2021)
exact sampling distribution of correlation coefficients and the bootstrap. 
When
the direction is known, proposed asymmetric (one-tail) tests have greater power. 
\end{abstract}

\section{Introduction}
\label{sec.intro}

A great deal of science focuses on understanding the
dependence between variables. Its quantification has a
long history starting with Galton-Pearson correlation
coefficient $r_{ij}$ from the 1890s and its cousins, including Spearman's $\rho$, Kendall's $\tau$, and Hoeffding's $D$.
Let $dep(X_i|X_j)$ measure the strength of dependence of $X_i$ on $X_j$
given a measurement $X_j$. Many measures of dependence try to satisfy
the symmetry postulate
by \cite{Renyi59}, which posits that the two strengths 
based on opposite conditioning are identical:
\begin{equation}
\label{eq.symm}
dep(X_i|X_j)\equiv dep(X_j|X_i).
\end{equation}
We regard the symmetry postulate akin to 
an avoidable dogma. The
following subsection explains why 
attempting to satisfy this symmetry
equation (\ref{eq.symm}) provides misleading measures of dependence
in practice.

\subsection{Four Examples of Asymmetric Dependence}
\label{sec.ex4}
A correct notion of dependence in nature or data is rarely (if ever)
symmetric. 
\begin{itemize}
	\item 
A newborn baby boy depends on his mother for his survival, 
but it is ludicrous to expect that his mother must exactly equally
depend on the boy for her survival, as implied by (\ref{eq.symm}). 
\item
Meteorologists know that the average daily high of December temperatures
in New York city is 44 degrees Fahrenheit and that this number 
depends on New York's latitude (40.7).  
The latitude is a
geographical given and does not depend on anything like
city temperatures. 
Symmetric dependence by (\ref{eq.symm}) between temperature and latitude
implies the ludicrous claim that latitude depends
on temperature with equal strength. 
\item For a third example, imagine a business person B with several shops.
B's 30\% earnings 
depend on the hours worked by a key employee in one shop. Now the symmetry  by (\ref{eq.symm}) means
that hours worked by the key employee
always depend on B's earnings, exactly 30\%.
%even if some earnings are from another shop.
\item
Our fourth example assumes $Y$ as complete data, but a subset of
$Y$ is unavailable.  The available subset $X$ is a proxy that depends on $Y$, but
the complete set $Y$ does not equally depend on its subset $X$.
\end{itemize}

These four examples are enough to convince the reader that the 
symmetry postulate is neither necessary nor sufficient
for real-world dependence.  However, it is interesting that the
unrealistic property (\ref{eq.symm}) is an old established
sacrosanct postulate from the 1950s, \cite{Renyi59}. 
Even in 2022, \cite{GPierre22}(``GM22''),
still adhere to the symmetry postulate (dogma)
by proposing an ingenious new definition of dependence to fit the model in (\ref{eq.symm}).  Actually,
measure of dependence satisfying (\ref{eq.symm}) can be
ludicrous some contexts analogous to the four examples above. 

\subsection{Sources of the Symmetry Dogma}
\label{sec.dogma}
What is the origin of symmetry dogma?
\begin{description}
	\item 
(i) The definitional and numerical
equality of covariances, $Cov(X_i, X_j) =Cov(X_j, X_i)$ may
have been the initial reason for the symmetry result.
\item 
(ii) 
In a bivariate linear regression $X_1=a+bX_2+\epsilon$, the strength of
dependence of $X_1$ on $X_2$ is clearly measured by
the coefficient of determination $R^2_{1|2}$.
If we consider a flipped linear regression,
$X_2=a^\prime+b^\prime X_1+\epsilon^\prime$, the strength of dependence is
$R^2_{2|1}$. The assumption of linearity makes the two strengths
equal to each other $R^2_{2|1}=R^2_{1|2}$. The equality of two
$R^2$ strengths supports the symmetry dogma. When we consider
the signed square roots of the two $R^2$ values, we have
a symmetric matrix of correlation coefficients,
$r_{ij}=r_{ji}$. These signed measures of dependence
further support the dogma.

The symmetry dogma 
depends on the harmless-looking
linearity assumption. 
Back in 1784, the German philosopher Kant said:
``Out of the crooked timber of humanity, no straight thing was ever made.'' Since
social sciences and medicine deal with human subjects, evidence supporting linearity and the implied symmetry dogma is missing.

\item 
(iii) Since all distances satisfy symmetry, it may have been
another reason behind Renyi's postulate.  
\item
(iv) The concept of statistical independence in probability
theory is symmetric.
It can be formulated in terms of the absence of any divergence
between a joint density and a product of two marginal densities,
\begin{equation}
	\label{eq.productOFmarginals}
	f(X_i X_j)= f(X_i)\, f(X_j).
\end{equation}
Since dependence is the opposite of independence,
it is tempting (but unhelpful) to impose symmetry on dependence as well.
\end{description}

\subsection{Statistical Independence in Contingency Tables}
\label{sec.cntn}
Two-way contingency tables  refer to tabulated data on a grand
total of $T$ observations distributed over an $r\times c$ matrix. There are two
categorical variables represented by
$r$ manifestations of row characteristics $R_i$ along
($i=1,2,\ldots r$) rows, and $c$ column characteristics $C_j$ along  
($j=1,2,\dots c$) columns. The body of the ($r\times c$) contingency table
has observed values $O_{ij}$ in a matrix cell located at
row  number $i$ and column number  $j$.
The joint probability $P(R_i,C_j)$ is simply
$O_{ij}/GT$, where GT denotes the grand total of the
tabulated numbers. The row margins of the contingency table have
row totals, $R_i=\Sigma_j O_{ij}$. The column margin
has column totals, $C_j=\Sigma_i O_{ij}$.
The marginal probabilities are $P(R_i)=R_i/GT$ and
$P(C_j)=C_j/GT$, which are also called unconditional
probabilities.

A conditional probability restricts the sample space
to the part of the Table which satisfies the specified 
condition, referring to a particular row $R_i$ or column $C_j$.
The direct computation of conditional probability
has the respective row or column sums in its denominator
instead of the grand total $GT$.
An equivalent calculation of conditional probability
defines $P(R_i|C_j)=P(R_i,C_j)/P(C_j)$, as 
a ratio
of the joint probability to the marginal probability
of the conditioning characteristic.
Analogous conditional probability 
conditioning on row characteristic is a ratio
of the same joint probability to the marginal probability of the
conditioning row,
$P(C_j|R_i)=P(R_i,C_j)/P(R_i)$.

In probability theory based on contingency tables,
the notion of statistical independence is studied by considering
 the following three criteria.
\begin{description}
\item
(a) $P(R_i,C_j)=P(R_i)P(C_j)$, joint probability equals the
product of marginals.
\item
(b) $P(R_i|C_j)=P(R_i)$, conditional probability equals unconditional or marginal probability.
\item
(c) $P(C_j|R_i)=P(C_j)$, the other conditional probability equals unconditional or marginal probability.
\end{description}
Note that criterion (a) is both necessary and sufficient
for independence. It is 
symmetric in that the joint probability
is the same even if we interchange the order and write it as
$P(C_j,R_i)$. However, data can satisfy (b) without satisfying (c), and
vice versa. Hence tests of independence typically rely
on the symmetric criterion (a).  However, dependence is the
opposite of independence and is generally asymmetric.
We find that using (b) and (c) helps avoid the
misleading symmetry postulate in the context of dependence.

It is customary to imagine a population of thousands of
contingency tables; the observed table is one realization
from that population. The null hypothesis  ($H_0$) is that row and
column characteristics are statistically independent. The sample
table may not exactly satisfy independence in the sense
of (a) to (c) above.  The testing problem is whether the
observed table of $O_{ij}$ values could have arisen from a
population where conditions (a) to (c) are satisfied.
That is,  $O_{ij}$ are numerically close enough to the
expected values $E_{ij}=R_iC_j$ obtained from the
cross-product of relevant marginal totals.

Pearson's Chi-square test statistic for ($H_0$) or independence of row effect
and column effect in a contingency table is
\begin{equation}
\label{eq.chisq}
\chi^2=(O_{ij}-E_{ij})^2/E_{ij},\quad df=(r-1)(c-1),
\end{equation} 
where $df$ denotes the degrees of freedom. 
Note that $\chi^2\in[0, \infty)]$  of (\ref{eq.chisq}) cannot
be computed unless we have contingency tables. 
Statisticians have long recognized that the magnitude of
$\chi^2$ 
cannot reliably 
measure the direction and strength of dependence.
This paper assumes that
a practitioner would want to know both the
general direction and strength of dependence.

\section{Symmetric Measures of Dependence}
\label{sec.fifj}

\cite{Granger04} (``Gr04'') is an important paper on formal testing for
statistical independence, especially for time series data. They
cite a survey by \cite{Tjo96} on the topic.  The novelty in Gr04
is in using
nonparametric nonlinear kernel densities in testing the equality (\ref{eq.productOFmarginals}) in their test of independence.
Unfortunately, Gr04 authors
adhere to the symmetry dogma by 
insisting that, similar to independence, a measure of dependence 
should be a symmetric distance-type `metric.'

\subsection{Dependence Measures and Entropy}
\label{sec.entropy}
Shannon defined information content in 1948 
as the amount of surprise in a
piece of information. His information is inversely proportional
to the probability of occurrence and applies to both discrete and
continuous random variables with probabilities defined by
a probability distribution $f(y)$. In the context of entropy,
let us use the fourth example of Section \ref{sec.ex4},
where $Y$ is the complete data and $X$ is a subset with some missing
observations. How does $X$ depend on $Y$? We develop a measure
of dependence using information theory, especially entropy.

Intuitively, entropy is our ignorance or the extent of disorder in
a system. 
The entropy $H(Y)$ is defined by the mathematical expectation of
the Shannon information or $E(-log f(y))$. 
And the conditional entropy of Y given X averaged over $X $is
\begin{equation}
\label{eq.hy|x}
H(Y|X)=-E[ E[log(f_{Y|X}(Y|X))|X]].
\end{equation}

The reduction in 
our ignorance
$H(Y)$ by knowing the proxy $X$ is $H(Y)-H(Y|X)$.
Mutual information $I_{mu}(X, Y)$ is defined as $H(x)+H(Y)-H(X,Y).$
It is symmetric since $I_{mu}(X, Y)=I_{mu}(Y, X)$.
The entropy-based measure of dependence is
\begin{equation}
	\label{eq.entrop}
	D(X;Y)=\frac{H(Y)-H(Y|X)}{H(Y)},	
\end{equation}
or proportional reduction
in entropy of $Y$ by knowing $X$.
\cite{Reimherr13} complain that 
(\ref{eq.entrop}) is not symmetric. By contrast, we view asymmetry
as a desirable property.

Neyman-Pearson showed that a way to distinguish between two distributions
$f(X)$ and $f(Y)$ for parameter $\theta$ 
is the difference between logs of their likelihood functions.
Shannon's relative entropy, also known as
Kullback–Leibler (KL) divergence, is the expected value of that difference,
\begin{equation}
\label{eq.KLdiv}
KLD=E(log f(\theta |X) - log f(\theta |Y)).
\end{equation}
It is easy to verify that KLD or relative entropy is not symmetric.

Gr04 authors state on page 650 that
``Shannon's relative entropy and almost all other
entropies fail to be `metric', as they violate either symmetry, or the triangularity rule, or both.''
We argue that asymmetry is an asset, not a liability, in light of four examples in
Section \ref{sec.ex4}.  Hence, we regard (\ref{eq.entrop}) 
or (\ref{eq.KLdiv}) as 
superior measures compared to the symmetric measure by Gr04.

D(X;Y) of (\ref{eq.entrop}) and KLD of (\ref{eq.KLdiv}) cannot be used directly on data vectors. They need frequency
distribution counts as input based on the grouping of data into bins (histogram class intervals). The choice of the number of bins is arbitrary, and D(X;Y) and KLD 
are sensitive to that choice. Hence, we do not recommend D(X;Y) or KLD as
a general-purpose measure of dependence.

\subsection{Dependence Measures and Fisher Information}
\label{sec.Fisher}
Fisher information measures the expected amount of information 
given by a random variable $Y$ about a parameter $\theta$ of interest.
Under Gaussian assumptions, the Fisher information is inversely proportional
to the variance.
\cite{Reimherr13} use the Fisher information to define a measure of
dependence.
Consider the estimation of a model parameter $\theta$ using 
$X$ as a proxy for unavailable $Y$.  That is,
$X$ is a subset of $Y$ with missing observations, as in the fourth
example of Section \ref{sec.ex4}. 
If the Fisher information for $\theta$ 
based on proxy $X$ is denoted by $\mathcal{I}_X(\theta)$, they
define a measure of dependence as:
\begin{equation}
	\label{eq.inforatio}
	D(X;Y)=\frac{\mathcal{I}_X(\theta)}{\mathcal{I}_Y(\theta)},	
\end{equation}
where $\mathcal{I}_X(\theta) \le \mathcal{I}_Y(\theta)$. 
Consider the special case where a proportion $p$
of the $Y$ data are missing in $X$ at completely random  locations.
Then, the measure of dependence (\ref{eq.inforatio}) equals $p$.
This measure
of dependence is almost acceptable because it is asymmetric, where subset $X$
being a proxy for $Y$ cannot be interchanged with $Y$,
except that $D(X;Y)$ of (\ref{eq.inforatio}) cannot be negative. Later,  we recommend
in Section \ref{sec.kernelReg} a more generally
applicable and intuitive measure of dependence.

%cite \cite{Tjo96} and others 
\subsection{Regression Dependence from Copulas}
\label{sec.copula}

Consider a two-dimensional joint (cumulative)
distribution function $F(X, Y)$ and marginal densities
$U=F_1(X)$ and $V=F_2(Y)$
obtained by probability integral transformations.
Sklar proved in 1959 that
a copula function $C(F_1, F_2)=F$ is unique if the components are
continuous. The copula function $C: [0, 1]^2\to [0,1]$ 
is subject to certain conditions forcing it to be a
bivariate uniform distribution function. It is
extended to the multivariate case to
describe the dependence structure of the
joint density. We have noted in section
\ref{sec.cntn} that a contingency table represents
the joint dependence structure of row and column 
characteristics. Copulas represent similar joint
dependence when row and column characteristics are
continuous variables rather than simple categories.

\cite{Dette13} (``DSS13'') define joint density as $F_{X,Y}$,
and conditional density of $Y$ given $X$ as  $F_{Y|X=x}$.
They
use uniform random variables  $U$ and $V$ to construct
copula $C$ as a
joint distribution function. The copula serves as their measure of dependence
based on the quality of regression-based prediction of $Y$ from $X$.
The flipped prediction of $X$ from $Y$ ignored by DSS13 is
considered in Section \ref{sec.kernelReg} in the sequel.

DSS13 assume
Lipschitz continuity, which implies that a copula is absolutely continuous
in each argument, so that it can be recovered from any of its partial derivatives by
integration. The conditional distribution $F_{V|U=u}$ is related to
the corresponding copula $C(X,Y)$ by
$F_{V|U=u}(v)=\partial_1 C_{X,Y}(u, v)$.

A symmetric measure of dependence proposed by DSS13 is denoted here
as
\begin{equation}
\label{eq.rcopula}
r_D(X,Y) =6 \int_0^1 \int_0^1 F_{V|U=u} (v)^2 dv du,
\end{equation}
where $r_D=0$ represents independence, and $r_D=1$ represents
almost sure functional dependence. DSS13 focus on $r_D$ filling
the intermediate range of the closed interval $[0,1]$ while ignoring the negative range $[-1, 0)$. Section \ref{sec.kernelReg} covers $[-1, 1]$, including
the negative range.
DSS13 rely on parametric copulas, making them
subject to identification problems,
as explained by \cite{Allen-jrfm22}.

The numerical computation of (\ref{eq.rcopula})
is involved since it requires
the estimation of the copula’s partial derivative.
DSS13 authors propose a kernel-based estimation method without
providing any ready-to-use computational tools for $r_D$.

Remark 3.7 in \cite{Beare10} states that symmetric copulas imply
time reversibility, which is unrealistic for economic and financial data. 
\cite{BOURI18} reject the symmetry dogma and note that their parametric
copula can capture tail dependence, which is important in a study
of financial markets. \cite{Allen-jrfm22} uses nonparametric
copula construction and asymmetric $R^*$ from \cite{VinodRao14}.
Allen's application to financial data shows that  cryptocurrencies do not help portfolio diversification.
%GM22 do not recommend copula methods since they cannot be used for %categorical
%variables.

\subsection{Hellinger Correlation $\eta$ as a Dependence Measure}
\label{sec.hell}
Now we turn to the recent GM22 paper mentioned earlier,
which proposes Hellinger correlation
$\eta$ as a new symmetric measure of the strength of dependence.  
They need to normalize to ensure that $\eta \in [0,1]$. GM22 denote
the normalized version as $\hat \eta$. 
GM22 authors explain why  dependence
axioms by \cite{Renyi59}  need
updating, while claiming that their $\eta$
satisfies all
updated axioms. Unfortunately, GM22
retain the symmetry axiom criticized in 
Section \ref{sec.ex4} above. An advantage of $\eta$ over
Pearson's $r_{ij}$ is that it incorporates some
nonlinearities.

Let $F_1$ and $F_2$ denote the known marginal distributions
of random variables $X_1$ and $X_2$, and let $F_{12}$ denote
their joint distribution. Now, GM22 authors ask readers
to imagine reconstructing
the joint distribution from the two marginals. The
un-intuitive (convoluted?) definition of the strength of dependence by GM22 
is the
size of the ``missing link'' in reconstructing the joint
from marginals. This definition allows GM22 to claim that
symmetry is ``unquestionable.''

GM22 authors define squared Hellinger distance $\mathcal{H}^2(X_1, X_2)$
as the missing link between $F_{12}$ and $F_1 F_2$. They approximate
a copula formulation of $\mathcal{H}^2$ using
the \cite{Bhatta43} affinity coefficient $\mathcal{B}$.
Let $C_{12}$ denote the copula of $(X_1, X_2)$, and $c_{12}$ 
denote its density.
The computation of $\hat\eta$ in the R package {\bf HellCor} uses
numerical integrals
$\mathcal{B}=\int\int \surd c_{12}$. Hellinger correlation
$\eta$ is
\begin{equation}
	\label{eq.eta}
	\eta=\frac{2}{\mathcal{B}^2}\{ \mathcal{B}^4+(4-3\mathcal{B}^4)^{1/2}-2  \}^{1/2}.
\end{equation} 
The Hellinger correlation is symmetric, $\eta(X_1,X_2)=\eta(X_2, X_1)$.

GM22 provide an R package {\bf HellCor} to compute $\hat\eta$ from data  as a  measure of dependence, and 
test the null
hypothesis of independence of two variables.

A direct and intuitive measure of
dependence in a regression framework is
the multiple correlation coefficient (of determination)
$R_{1|2}^2$.
It is symmetric because even if we flip $X_1$ and $X_2$,
the $R_{2|1}^2$ from linear regressions is exactly the same.
The reason for the equality of two flipped $R^2$ values,
$R^2_{(2|1)}=R^2_{(1|2)}$, is the
assumption of linearity of the two regressions.  
When we
relax linearity, the two $R^2$ values generally differ, 
$R_{2|1}^2 \ne R^2_{1|2}$. We argue
that quantitative researchers should reject the unrealistic
linearity assumption
in the presence of ready-to-use kernel regression
({\bf np} package) software. 

Kernel-based $R^2$ values of flipped regressions are rarely equal.
GM22  cite \cite{Janzing13} only to reject
such asymmetric dependence suggested by nonparametric regressions.

\section{Recommended Measures of Dependence}
\label{sec.kernelReg}
We have noted earlier that covariances satisfy
symmetry $Cov(X_i,X_j) =Cov(X_j,X_i)$. However, the sign
of symmetric covariances suggests the overall direction
of the dependence between the two variables. For example,
$Cov(X_i,X_j)<0$ means when $X_i$ goes up $X_j$ goes down, by and large.
Most of the symmetric measures of dependence discussed above fail to
provide this type of useful directional information except
Pearson's correlation coefficients $r_{ij}$. Hence, 
$r_{ij}$ has retained its popularity as a valuable measure
of dependence for over a century, despite assuming 
unrealistic linearity.

\cite{Zheng2012} reject the dogma by introducing nonsymmetric
generalized measures of correlation ($GMC\in [0,1]$), proving that
\begin{equation}
\label{eq.gmcs}
GMC(Y|X)\ne GMC(X|Y).
\end{equation}
Since GMCs fail to provide directional information in covariances
needed by practitioners,
\cite{VinodRao14} and \cite{Vinod15b} extend \cite{Zheng2012} to develop
a non-symmetric correlation matrix $R^*=\{ r^*_{ij}\}$,
where $r^*_{ij}\ne r^*_{ji}$, while providing an R package.
The R package {\bf generalCorr} uses kernel regressions to
overcome the linearity of $r_{ij}$ from the {\bf np} package
by Hayfield and Racine,
which can handle kernel regressions
among both continuous and discrete variables.
%%See also \cite{Vinod:22}.

Sometimes the research interest is focused on the strength of 
dependence, while the direction is ignored, perhaps because it is already
established.  In that case, one can use the R package {\bf generalCorr}
and the R function {\tt depMeas(,)}. It is defined
as appropriately signed 
larger of the two generalized correlations, or
\begin{equation}
	\label{eq.depMeas}
	depMeas(X_i, X_j)= sgn* max(|r^*(i|j)|,|r^*(j|i)|), 
\end{equation}
where $sgn$ is the sign of the covariance between
the two variables.

In general, both the strength and general
direction of quantitative dependence matter. Hence, we recommend two asymmetric measures
$r^*(X_i|X_j)$ and $r^*(X_j|X_i)$.  The {\bf generalCorr} package
functions for computing $R^*$ elements are {\tt rstar(x,y)} and {\tt gmcmtx0(mtx)}.
The latter converts a data matrix argument (mtx)  with $p$
columns, into a $p\times p$ asymmetric matrix $R^*$ of generalized
correlation coefficients.  Regarding the direction of dependence,
the convention is that the variable named in the column is the ``cause''
or the right-hand regressor, and the variable named along the row
is the response.  Thus the recommended measures from $R^*$ are
easy to compute. See an application to forecasting 
the stock market index of fear (VIX) and causal path determination in \cite{AllenHooper18}.

\subsection{Statistical Inference for Recommended Measures}
\label{sec.infer}
%\section{Statistical Inference for $r^*$}

We recommend the signed generalized correlation
coefficients $-1\le r^*_{ij}\ne r^*_{ji}\le 1$
from the $R^*$ matrix as the best dependence measure.
This is because they
do not adhere to the potentially misleading symmetry dogma while
measuring arbitrary nonlinear dependence dictated by the data.
An additional reason is its potential for more powerful 
(one-tail) inference,
discussed in this section.

The sign of each element of the $R^*$ matrix is based on the
sign of the covariance $Cov_{ij}=Cov(X_i, X_j)$. A two-tail
test of significance is appropriate only when $Cov_{ij}\approx 0$.
Otherwise, a one-tail test is appropriate.
Any one-tailed test provides greater power to detect an effect in one direction by not testing the effect in the other direction,
\cite{Kendallv2}, sections 22.24 and 22.28.

Since sample correlation coefficient $r_{ij}$
from a bivariate normal parent has a non-normal
distribution, Fisher
developed his famous z-transformation in the 1920s. He
proved 
that the following transformed statistic $r^T_{ij}$ is approximately normal
with a stable variance,
\begin{equation}
	\label{eq.fisherT}
	r^T_{ij}=(1/2) \quad log \frac{(1+r_{ij})}{(1-r_{ij})}
	\sim N(0, 1/n),
\end{equation}
provided $r_{ij}\ne 1$.
Recent work has developed the exact distribution of a correlation
coefficient.  It is now possible to directly compute
a confidence interval for any hypothesized value $\rho$ of the
population correlation coefficient.

Let $r$ be the empirical correlation of a random sample of size $n$ from
a bivariate normal parent. Theorem 1 of \cite{Taraldsen} generalized Fisher's famous z-transformation extended by C. R. Rao.
The exact density with $v=(n-1)>1$ is
\begin{eqnarray}
	\label{eq.tara}
	f(\rho| r,v)&=\frac{v(v-1)\Gamma(v-1))}{\surd(2\pi)\Gamma(v+0.5)}(1-r^2)^{\frac{v-1}{2 }}\,
	(1-\rho^2)^{\frac{v-2}{2}} (1-r \rho)^{ \frac{1-2v}{2} } \\\nonumber
	& \quad F(\frac{3}{2};-0.5; v+0.5; \frac{1+r\rho}{2}),
\end{eqnarray}
where F(.;.;.;.) denotes the Gaussian hypergeometric function, available in the
R package {\bf hypergeom} by R.K.S Hankin. The following R code
readily computes (\ref{eq.tara}) over a grid of 2001 $r$ values.
\begin{verbatim}
library(hypergeo); r=seq(-1,1,by=0.001)
Tarald=function(r,v,rho,cum){ #find quantile r given cum
Trm1=(v*(v-1)*gamma(v-1))/((sqrt(2*pi)*gamma(v+0.5)))
Trm2=(1-r^2)^((v-1)/2)
Trm2b=((1-rho^2)^((v-2)/2))*((1-rho*r)^((1-2*v)/2))
Trm3b=hypergeo(3/2,-1/2,(v+0.5),(1+r*rho)/2)
y0=Re(Trm1*Trm2*Trm2b*Trm3b) 
p=y0/sum(y0)
cup=cumsum(p)
loc=max(which(cup<cum))+1
return(r[loc])}
Tarald(r=seq(-1,1,by=0.001),v=11,rho=0,cum=0.05) #example
\end{verbatim}

\begin{figure}[h]
	\caption{Taraldsen's exact sampling density of  a
		correlation coefficient under the
		null  of $\rho=0$, solid line n=50, dashed line n=15}
	\label{fig.exactfr}
	\includegraphics[height=2.5in,width=4.5in]{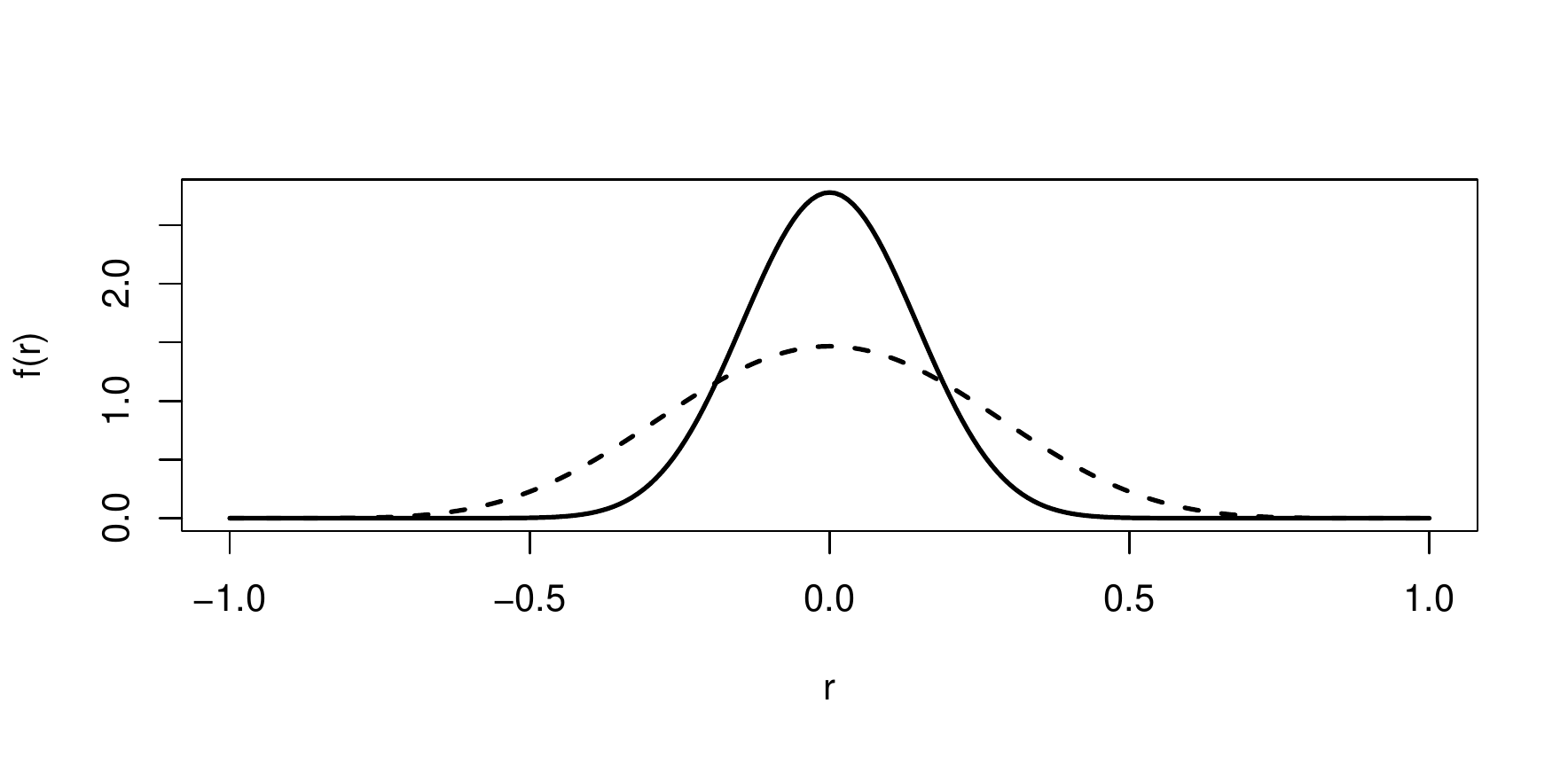}
\end{figure}

Assuming that the data come from a bivariate normal parent, 
the sampling distribution of
any
correlation coefficient is (\ref{eq.tara}). Hence,
the sampling distribution of unequal off-diagonal elements of
the matrix of generalized correlations $R^*$ also follows
(\ref{eq.tara}). When
we test the null hypothesis
$H_0: \rho=0$, the relevant sampling distribution 
is 
obtained by plugging $\rho=0$  in (\ref{eq.tara}) depicted in
Figure \ref{fig.exactfr}
for two selected sample sizes. Both distributions are centered at the null value $\rho=0$.

A two-tail (95\%, say) confidence interval is obtained
by using the 2.5\% and 97.5\% quantiles of the density.
If the observed correlation coefficient $r$ is inside the confidence interval,
we say that the observed $r$ is statistically insignificant,
as it could have arisen from a population 
where the null value $\rho=0$ holds.

\begin{figure}[h]
	\caption{Taraldsen's exact sampling density of correlation coefficient under the
		null  of $\rho=0.5$, solid line n=50, dashed line n=15}
	\label{fig.exactfr2}
	\includegraphics[height=2.5in,width=4.5in]{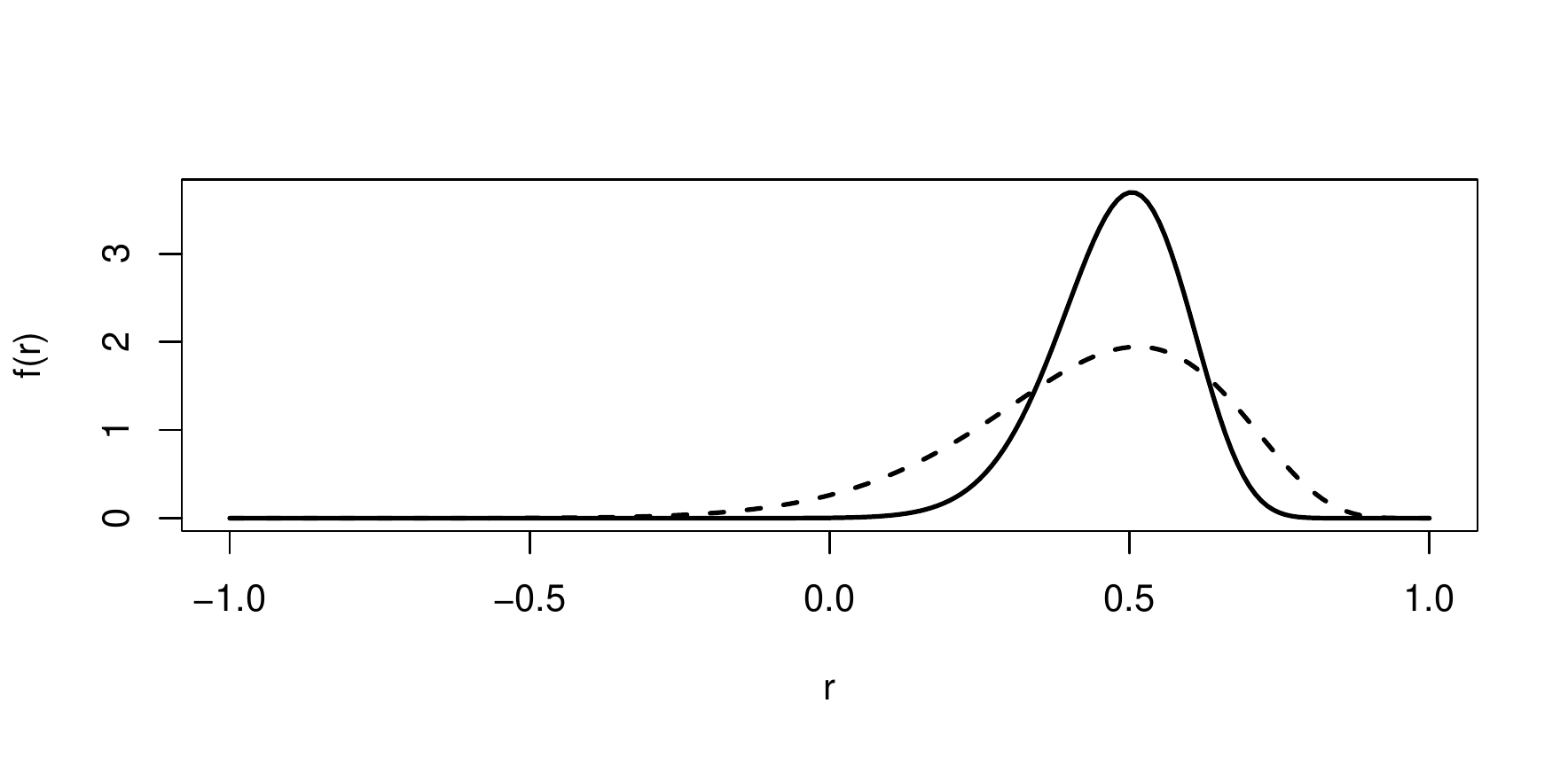}
\end{figure}

Similarly, one  can test the nonzero null hypothesis
$H_0: \rho=0.5$ using the equation 
obtained by plugging $\rho=0.5$ in (\ref{eq.tara})
depicted in Figure \ref{fig.exactfr2} .

Figures \ref{fig.exactfr} and \ref{fig.exactfr2} show that the
formula (\ref{eq.tara}) and our
numerical implementation are ready for practical use. 
These exact densities depend on the sample size and on the value
of the population correlation coefficient,  $-1 \le \rho \le 1$.
Given any hypothesized $\rho$ and sample size, a computer algorithm
readily computes the exact density, similar to Figures \ref{fig.exactfr} and \ref{fig.exactfr2}.  Suppose we want to help
typical practitioners who
want the tail areas useful for testing
the null hypothesis $\rho =0$. Then, we need to create a table of
a set of typical quantiles evaluated at certain cumulative probabilities and a corresponding selected set
of common sample sizes with a fixed  $\rho =0$.

Because of the complicated form of the density (\ref{eq.tara}),
it is not surprising that its (cumulative) distribution function $\int_{-1}^r f(\rho| r,v)$ by
analytical methods is not available in the literature. 
Hence, let us compute cumulative probabilities by
numerical integration defined as the rescaled area under the 
curve $f(r,v)$ for $\rho=0$. See Figure \ref{fig.exactfr}
for two choices of $v(=n-1)$ for sample sizes (n=50, 15). 
The cumulative probability becomes a sum of 
rescaled areas of small-width rectangles
whose heights are determined by the curve tracing $f(r,v)$.
The accuracy of numerical approximation to the area is obviously
better, the larger the number of rectangles.

We use a sequence of $r\in [-1, 1]$
created by the R command {\tt r=seq(-1,1, by =0.001)}, yielding 2001  rectangles. Denote
the height of $f(r,v)$ by $H_f=H_{f(r,v)}$. 
Now, the area between any two 
$r\in [-1, 1]$ limits,  say $r_{Lo}$ and $r_{Up}$ is a 
summation of areas (height times width=0.001) of all rectangles. Now, the cumulative probabilities in the range are
\begin{equation}
\label{eq.cumProb}
\Sigma_{r_{Lo}}^{r_{Up}} H_{f} / \Sigma_{-1}^{1} H_f,
\end{equation}
where the common width cancels, and where the denominator
$ \Sigma_{-1}^{1} H_f$ converts the
rectangle areas into probabilities.
More generally, we can use $f(\rho, r,v)$ for any $\rho \in [-1, 1]$.

Thus we have a numerical approximation 
to the exact (cumulative) distribution function
under the bivariate normality of the parent,
$$F(\rho,r,v)=\int_{-1}^r f(\rho| r,v).$$ The transform
from $f(.)$ to $F(.)$ is called the probability integral
transform, and its inverse $F^{-1}(c|\rho, v)$ gives 
relevant correlation coefficients $r$ as quantiles
for specified cumulative probability $c$ as the argument. 
A computer algorithm can readily find such quantiles.

The exact $F^{-1}(c|\rho,v)$ allows the construction of confidence intervals
based on quantiles for each $\rho$ and sample size. 
For example, a 95\% two-tail confidence
interval  uses the 2.5\% quantile $F^{-1}(c=0.025)$ as the lower limit,
and 97.5\% quantile $F^{-1}(c=0.975)$ as the upper limit.
These limits depend on hypothesized $\rho$ and sample size.
Since $\rho=0$ is a common null hypothesis for correlation
coefficients,
let us provide a table of $F^{-1}(c)$ quantiles for
eleven sample sizes (listed in row names) and
eight cumulative probabilities 
listed in column titles of Table \ref{tab.1}.

The p-values in statistical inference
are defined as the probability of
observing the random variable
(correlation coefficient) as extreme or more extreme than
the observed value of the correlation coefficient $r$
for a given null
value $\rho=0$.  
Any one-tail p-values based on
$f(\rho| r,v)$ of (\ref{eq.tara}) for arbitrary nonzero ``null'' values of $\rho$ can be similarly computed by numerical integration
defined as the area under the curve.
Some code for
R functions {\tt Tarald(.)}, and {\tt pTarald(.)} is included in 
Sections \ref{sec.infer} and \ref{sec.examp}, respectively.

% latex table generated in R 4.2.1 by xtable 1.8-4 package
% Thu Nov 10 19:06:49 2022
\begin{table}[ht]
\centering
\caption{Correlation coefficients as quantiles
evaluated at specified cumulative probabilities (c=.) using Taraldsen's exact sampling distribution for various sample sizes
assuming $\rho=0$}
\label{tab.1}
	\begin{tabular}{rrrrrrrrr}
		\hline
		&        & c= &              &    &      &       & c= &  \\ 
		& c=0.01 & 0.025 & c=0.05 & c=0.1 & c=0.9 & c=0.95 & 0.975 & c=0.99 \\ 
		\hline
		n=5 & -0.83 & -0.75 & -0.67 & -0.55 & 0.55 & 0.67 & 0.75 & 0.83 \\ 
		n=10 & -0.66 & -0.58 & -0.50 & -0.40 & 0.40 & 0.50 & 0.58 & 0.66 \\ 
		n=15 & -0.56 & -0.48 & -0.41 & -0.33 & 0.33 & 0.41 & 0.48 & 0.56 \\ 
		n=20 & -0.49 & -0.42 & -0.36 & -0.28 & 0.28 & 0.36 & 0.42 & 0.49 \\ 
		n=25 & -0.44 & -0.38 & -0.32 & -0.26 & 0.26 & 0.32 & 0.38 & 0.44 \\ 
		n=30 & -0.41 & -0.35 & -0.30 & -0.23 & 0.23 & 0.30 & 0.35 & 0.41 \\ 
		n=40 & -0.36 & -0.30 & -0.26 & -0.20 & 0.20 & 0.26 & 0.30 & 0.36 \\ 
		n=70 & -0.27 & -0.23 & -0.20 & -0.15 & 0.15 & 0.20 & 0.23 & 0.27 \\ 
		n=90 & -0.24 & -0.20 & -0.17 & -0.14 & 0.14 & 0.17 & 0.20 & 0.24 \\ 
		n=100 & -0.23 & -0.20 & -0.16 & -0.13 & 0.13 & 0.16 & 0.20 & 0.23 \\ 
		n=150 & -0.19 & -0.16 & -0.13 & -0.10 & 0.10 & 0.13 & 0.16 & 0.19 \\ 
		\hline
	\end{tabular}
\end{table}

For the convenience of practitioners, we explain how
to use the cumulative probabilities
in Table 1 in the context of testing the null hypothesis $\rho=0$. The Table
confirms that the distribution is symmetric around $\rho=0$ as in
Figure \ref{fig.exactfr}. Let us consider some examples.
If n=100, the critical value from Table 1 for a one-tail 95\%
test is 0.16 (line n=100, column c=0.95). Let the observed
positive r be 0.3. Since $r$ exceeds the critical value, 
$(r>0.16$), we reject $\rho=0$.
If n=25, the critical value for a 5\% left tail in Table 1 is
$-0.32$. If the observed $r=-0.44$, is less than
the critical value $-0.32$ it falls in the left tail,
and we reject $\rho=0$ to conclude that it is significantly negative.

Table 1 can be used for constructing a few two-tail 95\% confidence intervals as follows.
If the sample size is 30, we see along the row n=30, and column
c=0.025 gives $-0.35$ as the lower limit, and column c=0.975 gives
$0.35$ as the upper limit. In other words, for n=30, any correlation
coefficient smaller than 0.35 in absolute value is statistically
insignificant.

If the standard bivariate normality assumption is not believed, one can
use the maximum entropy bootstrap (R package {\bf meboot}) designed for
dependent data.  A bootstrap application creates a large number 
$J=999$, say, versions
of data ($X_{i\ell}, X_{j\ell}$) for $\ell=1, \ldots J$. Each version
yields  $r^*(i|j;\ell), r^*(j|i;\ell)$ values. A large set of
$J$ replicates of these correlations give
a numerical approximation to the sampling distribution of these
correlations. Note that such a bootstrap sampling distribution is data-driven.
It does not assume bivariate normality needed 
for the construction of Table 1 based on (\ref{eq.tara}).

Sorting the replicated $r^*(i|j;\ell), r^*(j|i;\ell)$ values from the smallest to
the largest, one gets their ``order statistics'' denoted
upon inserting parentheses by replacing
$\ell$ by $(\ell)$. Now a
left-tail 95\% confidence interval for $r^*(i|j)$
leaves a 5\% probability mass
in the left tail. The interval is approximated by the order statistics as
$[r^*(i|j;(50)), 1]$. If the hypothesized $\rho=0$ is
inside the one-tail interval, one fails to reject (accepts) the null hypothesis 
$H_0: \rho=0$.

We conclude this section by noting that recommended measures of
dependence based on the $R^*$ matrix and their formal inference
are easy to implement. The tabulation of Taraldsen's exact sampling distribution
of correlation coefficients in Table 1 is new and should be of 
broader applicability. It is an improvement over standard significance
tests of correlation coefficients based on Fisher's z-transform. 
The next section illustrates with examples the use of Table 1,
newer dependence measures, and
other inference tools.

\section{Dependence Measure Examples \& Tests}
\label{sec.examp}
This section considers some examples of dependence measures.
Our first example deals with fuel economy in automobile design.
R software comes with `mtcars' data on 
ten aspects of automobile design and performance for 32 automobiles.
We consider two design features for illustration, 
miles per gallon $mpg$ and horsepower  $hp$.
\cite{VinodRao14} reports the Pearson correlation coefficient
$r(mpg,hp)=-0.78$ in his Figure 2. The negative sign correctly
shows that one gets reduced $mpg$ when a car
has larger horsepower $hp.$  Table 2 in \cite{VinodRao14} reports two generalized correlation coefficients obtained by using
kernel regressions as
$r^*(mpg|hp)=-0.938$ and $r^*(hp|mpg)=-0.853$.

One can interpret these $r^*(X_i|X_j)$ values as signed strengths of
dependence of $X_i$ on the conditioning variable $X_j$. 
The strengths are asymmetric,
$|r^*(mpg|hp)|>|r^*(hp|mpg)|$,
the absolute
value of a 
dependence strength using both generalized correlation coefficients
is larger than the dependence strength suggested under linearity.
Thus Pearson's correlation coefficient can
underestimate dependence  by assuming linearity.

 For the
`mtcars' data depMeas based on (\ref{eq.depMeas}) is
$-0.938$.
Now consider Table 1 for n=30 row and c=0.05 for a one-tail
critical value $-0.30$. The observed correlation $-0.938$ is
obviously in the left tail (rejection region) of the exact sampling distribution
of the correlation coefficient. Thus the negative dependence
of fuel economy (mpg) on the car's horsepower is
statistically significant. 
We re-confirm the significance by computing the one-tail p-value 
(= 1e-16) using an R function {\tt pTarald(.)}.  
Our R code for p-values from
Taraldsen's exact density of correlation coefficients is 
given next.
\begin{verbatim}
pTarald=function(r,n,rho,obsr){
v=n-1
if(v<=164)  Trm1=(v*(v-1)*gamma(v-1))/((sqrt(2*pi)*gamma(v+0.5)))
if(v>164)  Trm1=(164*(163)*gamma(163))/((sqrt(2*pi)*gamma(163.5)))
Trm2=(1-r^2)^((v-1)/2)
if(rho!=0)  Trm2b=((1-rho^2)^((v-2)/2))*((1-rho*r)^((1-2*v)/2))
if(rho==0)  Trm2b=1
Trm3b=Re(hypergeo(3/2,-1/2,(v+0.5),(1+r*rho)/2))
y0=Re(Trm1*Trm2*Trm2b*Trm3b) 
p=y0/sum(y0)
cup=cumsum(p)
loc=max(which(r<obsr))+1
if(obsr<0) ans=cup[loc]
if(obsr>=0) ans=1-cup[loc]
return(ans)}
pTarald(r=seq(-1,1,by=0.001),n=32,rho=0,obsr=-0.938)
\end{verbatim}
The first term (Trm1) in the R function computing the p-values
involves a ratio of two gamma (factorial) functions
appearing in (\ref{eq.tara}). For $n>164$ each gamma becomes infinitely
large, and Trm1 becomes `NaN' or not a number. Our code
winsorizes large $n$ values.

Since mtcars data has n=32, and the observed generalized correlation is $r^*=0.938$,
we use the command on the last line of the code to get the p-value
of 1e-16, or extremely small, suggesting statistical significance.
If we use 
the same automobile data using GM22's R package called
{\bf HellCor} we find that
$\eta=0.845>0$, giving no hint
that mpg and hp are negatively related.
If we compare numerical magnitudes, we have
$\eta>|r(mpg,hp)|=0.78$. Since $\eta$ exceeds Pearson's
correlation in absolute value,
$\eta$ is seen to incorporate nonlinear dependence. 
However, $\eta=0.78$ may be 
an underestimation of the absolute value of depMeas=$-0.938$. We fear that either $\eta$ may be failing to incorporate
some nonlinear dependence or is paying an unknown penalty for adhering to
the symmetry dogma.

\subsection{Further Real-Data Applications in GM22}
\label{sec.fishbirds}
GM22 illustrate the
Hellinger correlation measure of dependence using two sets of data
where the Pearson correlation is statistically insignificant,
yet their Hellinger correlation is significant.
Their first data set
refers to the population of
seabirds and coral reef fish residing around n = 12 islands 
in the British Indian Ocean Territory of Chagos Archipelago.
Ecologists and other scientists cited by GM22 have determined
that fish and seabirds have an
ecologically symbiotic relationship. The seabirds create an
extra nutrient supply to help algae. Since fish primarily
feed on those algae, the two variables should have a significantly positive dependence.

GM22 begin with the low Pearson correlation $r(fish, seabirds)=0.374$
and a 95\% confidence interval  $[-0.2548,  0.7803]$
that contains a zero, suggesting
no significant dependence. The p-value using {\tt pTarald(..,obsr=0.374)}
is 0.0935, which exceeds the benchmark of 0.05, confirming statistical insignificance.
The wide confidence interval, which
includes zero, is partly due to the small sample size (n=12).

Our Table 1 with the exact distribution of correlations suggests that
when $n=10$, more conservative than the correct n=12, the exact two-tail 95\% confidence interval (leaving 2.5\% probability mass in 
both tails) 
also has
a wide range $[-0.58, 0.58]$, which includes zero. Assuming the direction is known, a one-tail interval
with 5\% in the right tail (n=10) value is 0.50.  That is, only when the
observed correlation is larger than 0.50 it is significantly
positive (assuming a bivariate normal parent density).

\begin{figure}[h]
	\caption{Marginal densities of fish and
seabirds data are skewed, not Normal}
	\label{fig.density}
\includegraphics[height=2.5in,width=4.5in]{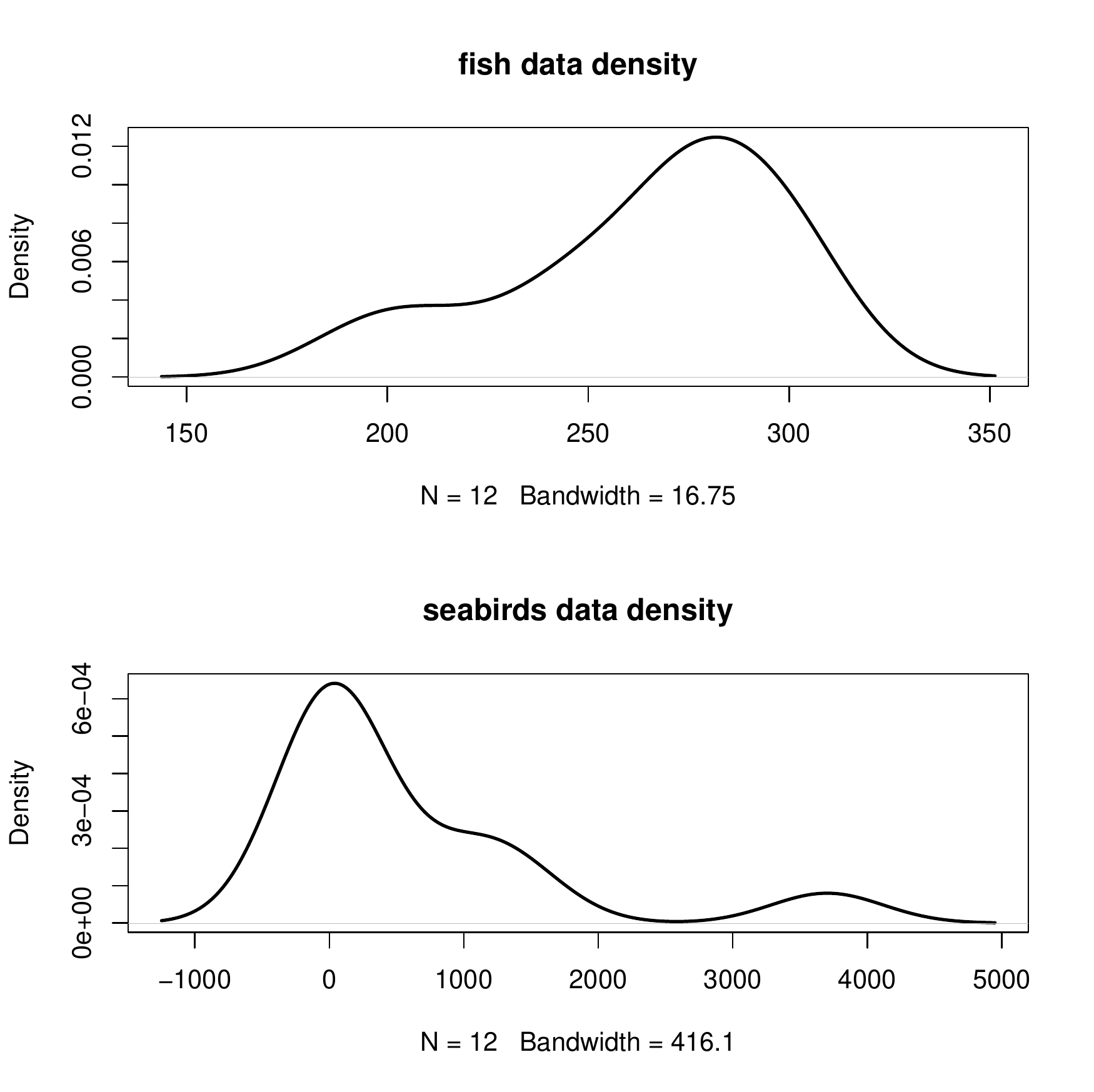}
\end{figure}

GM22 find that their Hellinger correlation $\eta$ needs to be
normalized to ensure that $\eta \in [0,1]$, because their
estimate of $\mathcal{B}$ can exceed unity. 
They denote
the normalized version as $\hat \eta$ and claim an
easier interpretation of $\hat\eta$  on the ``familiar Pearson scale,''
though Pearson's $r_{ij}\in [-1, 1]$ scale admits negative values.
GM22 employ considerable ingenuity to achieve the positive range
[0, 1] described in their Section 5.3. They 
state on page 650 that their range normalization ``comes at the price of
a lower power when it comes to test for independence.''

Using the population of
seabirds and coral reef fish residing around n = 12 islands,  
GM22 report the estimate
$\hat \eta$(fish, seabirds)=0.744. 
If one assumes bivariate normal parent distribution
and uses Taraldsen's exact density from Table 1,
$\hat\eta$(fish, seabirds)$=0.744 >0.50$ suggests statistical significance.
The p-value using {\tt pTarald(..,obsr=0.744)}
is 0.0027, which is smaller than the benchmark 0.05, confirming significance.

In light of Figure \ref{fig.density}, it is
unrealistic to
assume that the data come from a bivariate normal parent distribution.
Hence the evidence showing a significantly positive correlation
between fish and seabirds based on Taraldsen's exact density
is suspect.  Accordingly,
GM22 report a bootstrap p-value of
$0.045<0.05$ as their evidence. Since this p-value is too close to 0.05, we check for unintended p-hacking.
%All bootstrap estimates are sensitive to random seeds.
When one runs their {\tt HellCor(.)} function with 
{\tt set.seed(99)} and default settings, the bootstrap p-value becomes $0.0513>0.05$, which
exceeds the benchmark suggesting insignificant $\hat\eta (fish,seabirds).$
Then, GM22's positive Hellinger correlation estimate of $\hat\eta$= 0.744 is not
statistically significant at the usual 95\% level.
Thus, the Hellinger correlation fails to be strongly
superior to Pearson's correlation $r$ because $\hat\eta$
is also insignificantly positive.

\begin{figure}[h]
\caption{Bootstrap density of generalized correlation coefficient
		r*(seabirds, fish).}
\label{fig.bootdensity}
\includegraphics[height=2.5in,width=4.5in]{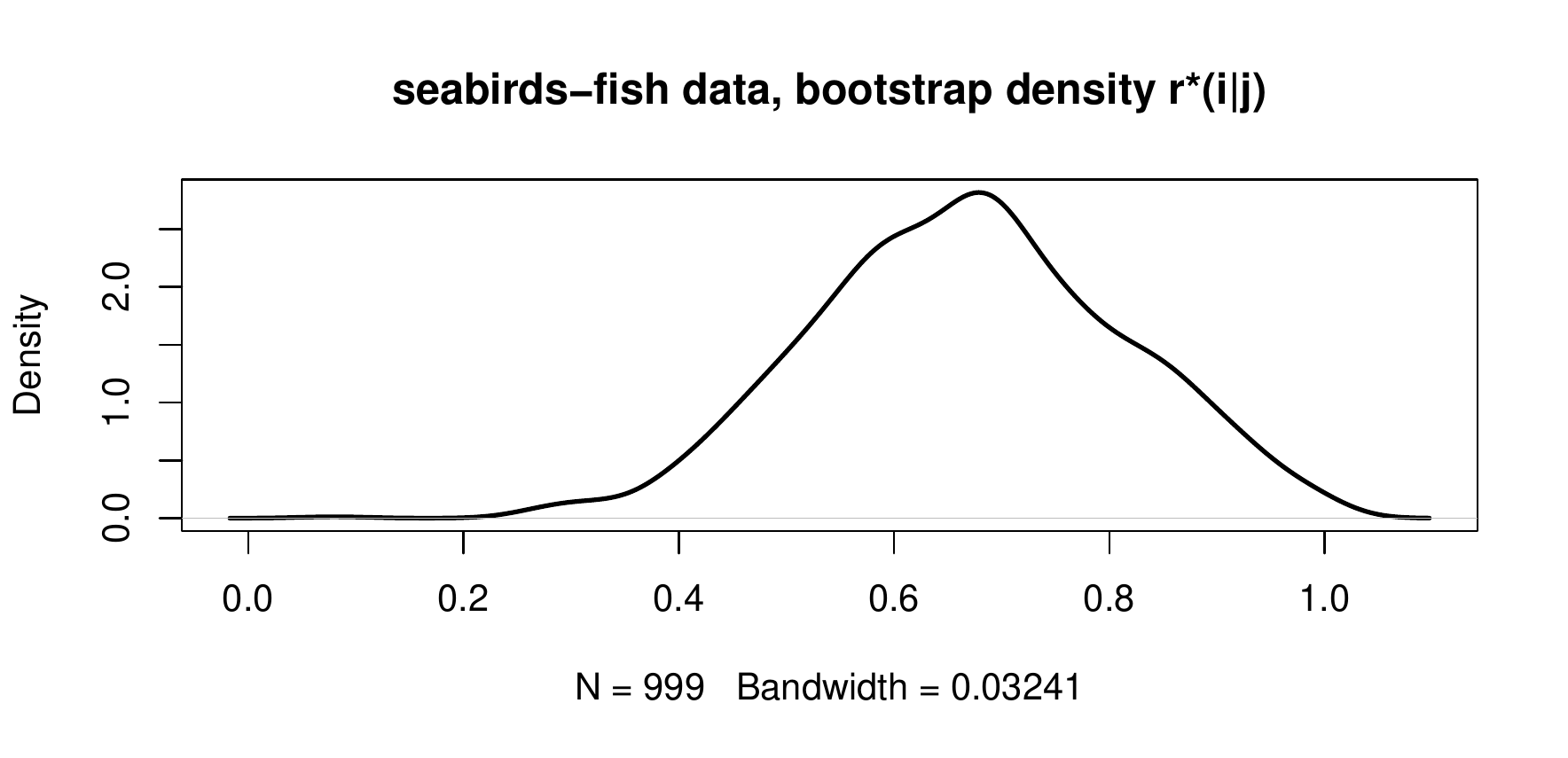}
\end{figure}

Now, let us compare $\hat \eta$ with off-diagonal
elements of the generalized correlation matrix $R^*$ recommended here.
Our {\tt gmcmtx0(cbind(fish,seabirds))} suggests the ``causal'' direction
(seabirds$\to$ fish), to be also positive, $r^*(fish| \break seabirds) =0.6687$.
The p-value using {\tt pTarald(..,obsr=0.6687)}
is 0.0086, which is smaller than the benchmark 0.05, confirming significance. There is no suspicion of
p-hacking here.
A 95\% bootstrap two-tail confidence interval using the {\bf meboot} R package is
[0.3898, 0.9373]. A one-tail interval is [0.4394, 1], which includes
the observed 0.6687 with a p-value of zero. See Figure \ref{fig.bootdensity},
where almost the entire density has positive support.
Note that the interval does not include a zero, suggesting
significant positive dependence consistent with what the ecologists expect.
The lower limit of our {\bf meboot} confidence interval is not close to zero.
More importantly, our $R^*$ generalized correlation coefficients do not impose symmetric dependence, revealing
sign information borrowed from the covariance, absent in the Hellinger correlation.

The second example in GM22 has
the number of births ($X_1$) and deaths ($X_2$) per year per 1000 individuals in n=229 countries in 2020.
A data scatterplot in their Figure 7 displays a C-shaped nonlinear 
relation. Pearson's correlation $r_{12}=-0.13$ is negative and insignificant at level $\alpha = 0.05$. This is based on 
a two-tail
traditional Fisher approximation to the
sampling distribution of a correlation coefficient. It is reversed by
our more powerful one-tail p-value using Taraldsen's exact sampling distribution.
Our {\tt pTarald(..,n=229, obsr=-0.13)}
is $(0.0246<0.05)$,  implying a statistically significant negative
correlation. 
On the other hand, GM22 estimate $\hat\eta = 0.69$ with
a two-tail 95\% bootstrap
confidence interval [0.474, 0.746], hiding important
information about the negative direction of dependence.
Since zero is outside
the confidence interval, GM22 claim that they have 
correctly overcome
an apparently incorrect inference based on traditional methods.
We have shown that traditional inference was incorrect because
more accurate Taraldsen distribution was not used.

\begin{figure}[h]
\caption{Bootstrap density of generalized correlation coefficient
		r*(birth, death).}
\label{fig.birthdensity}
\includegraphics[height=2.5in,width=4.5in]{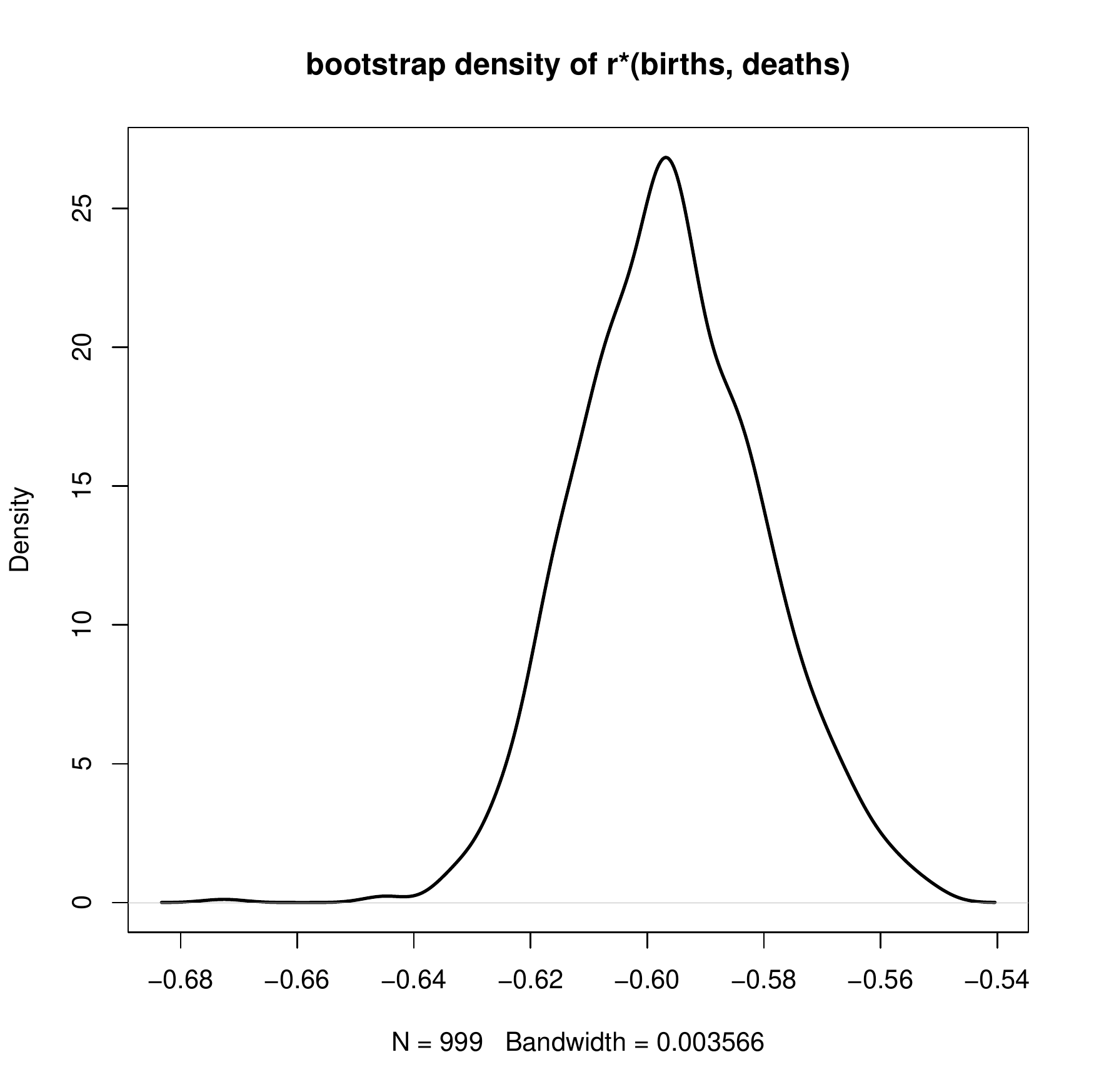}
\end{figure}

Our {\tt gmcmtx0(cbind(birth,death))} estimates that
$r^*(death|birth)$ is $=-0.6083$. A one-tail 95\% confidence
interval using the maximum entropy bootstrap (R package {\bf meboot}) 
is [-1, -0.5693]. 
 A somewhat less powerful
two-tail interval [$-0.6251, -0.5641$]  is also entirely negative.
The null hypothesis states that the true unknown $r^*$ is zero.
Since
our random interval excludes zero, the dependence is significantly 
negative. The p-value is zero in Figure \ref{fig.birthdensity},
since almost the entire density has negative support.
A larger birth rate significantly leads to a lower death rate in
229 countries in 2020.

In summary, the two examples used by GM22 to sell their Hellinger correlation
have a discernible advantage over Pearson's $r_{ij}$, but not
over our generalized correlation $R^*$. The examples confirm
four shortcomings of
Hellinger's correlation $\hat\eta$ over $R^*$. (a) It imposes an unrealistic symmetry assumption. (b) It provides no information about the direction of
dependence. (c) It forces the use of less powerful two-tail 
confidence intervals. (d) It is currently not implemented for discrete variables.

\section{Final Remarks} 
\label{sec.fin}
Many scientists are interested in measuring the directions and
strengths of dependence between variables.
This paper surveys quantitative measures of dependence between two variables.
We use four real-world examples in Section \ref{sec.ex4} to
show that any symmetric measure of dependence alleging
equal strength in both directions is unacceptable.
Yet, the majority of statistical dependence extant in the literature
adheres to the symmetry dogma.  A 2022 paper (GM22) develops
an intrinsically flawed
symmetric measure of dependence while proposing Hellinger correlations.

We show that off-diagonal elements
of the asymmetric $R^*$ matrix of generalized correlation coefficients
provide an intuitively sensible measure of dependence after incorporating
nonlinear and nonparametric relations among the variables involved.
The R package {\bf generalCorr} makes it easy to
implement our proposal. Its six vignettes provide ample illustrations
of the theory and applications of $R^*$.

We discuss statistical inference for elements of the $R^*$ matrix, providing
a new Table 1 of quantiles of \cite{Taraldsen} exact density
of a correlation coefficient for eleven typical sample sizes and eight
cumulative probabilities.
We illustrate with two data sets used by
GM22 to support their Hellinger correlation. 
Directional information is uniquely provided by our asymmetric measure
of dependence in the form of 
generalized
correlation coefficients 
{$\{r^*(i | j)\}$}. It
allows the researcher to achieve somewhat better qualitative
results and more powerful one-tail tests
compared to symmetric measures of dependence in the literature.

We claim that 
one-tail p-values of the
Taraldsen's density can overcome the inaccuracy of the traditional Pearson
correlation inference based on
Fisher's z-transform. We illustrate the claim using GM22's second
example where Pearson correlation $r(birth, death)$ is shown to be 
significantly negative using Taraldsen's density. Hence,
the complicated Hellinger correlation inference is not really 
needed to achieve correct significance. Interestingly, both
hand-picked examples designed to show the superiority of GM22's $\hat\eta$ over $r_{ij}$
show the merit of our proposal based on $R^*$ over $\hat\eta$.

Almost every issue of every quantitative journal refers to correlation
coefficients at least once, underlining its importance in measuring dependence.
We hope that $R^*$ and our implementation of Taraldsen's exact
sampling distribution of correlation coefficients receive further attention
and development.

\bibliographystyle{asa}
% %\bibliographystyle{Chicago}
\bibliography{ref12}

\end{document}